\documentclass[letter]{aa}  

\usepackage{graphicx}
\usepackage{txfonts}
\usepackage[citecolor=blue, linkcolor=blue, urlcolor = black, colorlinks = true]{hyperref}
\usepackage{ulem}
\usepackage{multirow}

\usepackage{graphicx}	
\usepackage{amsmath}	
\usepackage{amssymb}	
\usepackage{xspace}

\usepackage{siunitx}
\usepackage{comment}
\usepackage{textcomp}
\usepackage{bm}
\usepackage{lscape}
\usepackage{xcolor}
\newcommand{\eps}{\varepsilon}
\newcommand{\dt}[1]{\frac{\partial  #1}{\partial t}}
\newcommand{\vv}{\bold{v}}
\newcommand{\vn}{\vec{n}}
\newcommand{\calN}{ {\cal N}}

\newcommand{\Msun}{\,M$_{\odot}$\xspace}
\newcommand{\dMsun}{\,M$_{\odot}$\xspace}

\newcommand{\xc}{$X_{\rm c}$\xspace}
\newcommand{\xcx}{$X_{\rm c}/X_{\rm ini}$\xspace}

\newcommand{\percent}{~per~cent\xspace}

\newcommand{\omegai}{$(\Omega_{\rm eq} / \Omega_{\rm c})_{\rm i}$\xspace}

\newcommand{\edit}[1]{#1}

\newcommand{\mesa}{\texttt{MESA}\xspace}
\newcommand{\ester}{\texttt{ESTER}\xspace}

\defcitealias{Burssens2023}{B23}

\begin{document}

   \title{The first two-dimensional stellar structure and evolution models of rotating stars }
   \subtitle{Calibration to $\beta$ Cephei pulsator HD192575}

   \author{J.S.G. Mombarg\inst{1}
   \and 
   M. Rieutord\inst{1}
   \and
   F. Espinosa Lara\inst{2}
          }
   \institute{IRAP, Universit\'e de Toulouse, CNRS, UPS, CNES, 14 avenue \'Edouard Belin, F-31400 Toulouse, France
   \and 
   Space Research Group, University of Alcal\'a, 28871 Alcal\'a de Henares, Spain\\
              \email{jmombarg@irap.omp.eu}
        }

   \date{Received July 13 2023; accepted August 14 2023}
\titlerunning{2-D stellar evolution}
\authorrunning{Mombarg et al.}

 
  \abstract
   {Rotation is a key ingredient in the theory of stellar structure and evolution. Until now, stellar evolution codes operate in a one-dimensional framework for which the validity domain in regards to the rotation rate is not well understood.}
   {This letter aims at presenting the first results of self-consistent stellar models in two spatial dimensions that  compute the time evolution of a star and its rotation rate along the main sequence together with a comparison to observations.}
   {We make use of an extended version of the \ester code that solves the stellar structure of a rotating star in two dimensions with time evolution, including chemical evolution, and an implementation of rotational mixing. We have computed evolution tracks for a 12\dMsun model, once for an initial rotation rate equal to 15\percent of the \edit{critical} frequency, and once for 50\percent.   }
   {We first show that our model initially rotating at 15\percent of the \edit{critical} frequency is able to reproduce all the observations of the $\beta$ Cephei star HD 192575 recently studied by Burssens et al. with asteroseismology. Beyond the classical surface parameters like effective temperature or luminosity, our model also reproduces the core mass along with the rotation rate of the core and envelope at the estimated age of the star. This particular model also shows that the meridional circulation has a negligible influence on the transport of chemical elements, like nitrogen, for which the abundance may be increased at the stellar surface. Furthermore, it shows that in the late main sequence, nuclear evolution is faster than the relaxation time needed to reach a steady state of the star angular momentum distribution. }
   {We have demonstrated that we have successfully taken the new step towards two-dimensional evolutionary modelling of rotating stars. It opens new perspectives on the understanding of the dynamics of fast rotating stars and on the way rotation impacts stellar evolution.}


   \maketitle
%

\section{Introduction}
Over the next decades, stellar structure and evolution (SSE) theory will start advancing from the one-dimensional models towards solving the stellar structure equations in three spatial dimensions. So far, all SSE codes rely on the approximation of spherical symmetry as this greatly simplifies the numerics.    
But asteroseismic and interferometric studies of stars on the main sequence have revealed that a significant fraction of stars rotate with velocities where the approximation of spherical symmetry is not justified. 
\edit{Many} SSE codes that account for rotational effects are based on the pioneering works by \cite{Zahn1992} and \cite{Chaboyer1992}, and sequential works \citep[e.g.][]{Talon1998, Palacios2003, Mathis2004}, which provide expressions for the efficiency of chemical mixing due to both shear-driven turbulence and advective transport \citep[cf.][]{Ekstrom2012}. As this paradigm is constructed for one dimension, several assumptions have been made regarding the relative strengths of the horizontal and vertical diffusion, and the validity domain of these assumptions is currently unknown, except that they should apply in the limit of slow rotation.

Attempts at solving the stellar structure of rotating stars in more than one dimension already started in the sixties, but it has been only recently that the stellar structure can be computed in a self-consistent manner in two dimensions with the \ester\footnote{\url{https://github.com/ester-project/ester}} code \citep{Espinosa2013, Rieutord2016}. These models have been restricted to a steady-state, where the hydrogen-mass fraction profile is modelled as a step function, namely a constant value in the convective core and a constant value in the radiative envelope \cite[see e.g.][]{  Gagnier2019a, Gagnier2019b, Bouchaud2020, Howarth2023}. Furthermore, \ester has been designed for early-type stars and the applicability is currently limited to stars with a convective core and radiative envelope.

In this letter, we present the results of adding, for the first time ever, the temporal dimension to the two-dimensional steady-state models, by solving in addition the equation of chemical evolution. The overall aim of the present work is to demonstrate the capabilities of the state-of-the-art 2-D evolution \ester models of a rotating massive star, and present the predictions for the rotation and chemical profiles. This includes fully solving the large-scale velocity field inside the star. We successfully performed two-dimensional evolution of a slow and a moderate rotator with a mass of 12~\Msun, from zero-age main sequence (ZAMS) up to near terminal-age main sequence. For the predicted rotation profile, this work also aims at making a comparison with observations. In Sect.~\ref{sec:ester}, we describe the fundamentals of the \ester code and the improvements made. In Sect.~\ref{sec:astero}, we compare the predictions of the 2-D models, especially the predicted rotation profile, with asteroseismic measurements of a massive star. We show that the predictions of the two-dimensional model are consistent with the observations. In Sect.~\ref{sec:mixing}, we discuss the contribution of meridional circulation to nitrogen enrichment as predicted by 2-D models. Finally, we conclude in Sect.~\ref{sec:conclusions}.

\section{ESTER models} \label{sec:ester}

In this work, we make use of new \ester models where time evolution has been implemented. We recall that \ester models are two-dimensional models of rotating stars that include the centrifugal distortion of the star and the associated baroclinic flows. These models therefore predict the structure (pressure, density and temperature distributions) and the associated large-scale flows, namely the differential rotation and meridional circulations. The first version of the \ester code, which computes the steady states of early-type star models \citep{Espinosa2013}, has now been completed with a time evolutive version that includes the unsteady terms needed to  follow the thermal and nuclear evolution of a star. The code solves the set of equations given in Appendix~\ref{ap:eq}, still assuming the axisymmetry of the star. Additionally, in this new version, the evolution of the mass fraction of hydrogen is solved,
  \begin{equation} \label{eq:X_evol}
  \dt{X}+\vv\cdot\nabla X = \frac{1}{\rho}\nabla(\rho [D]\nabla X) + \dot X_{\rm nuc},
  \end{equation}
where $\rho$ is the density, $\vv$ is the meridional velocity, and $[D]$ is a tensor constructed from the horizontal and vertical chemical diffusion coefficients. 

As for the steady-state version we are using OPAL for opacities and equation of state \citep{Rogers1996}. Nuclear energy production is modelled via a simple law for the CNO cycle (see appendix) from which we derive $\dot X_{\rm nuc}$. 

The spatial discretization is based on a spectral element method where spectral elements are spheroidal shells bounded by isobars \cite[][]{Rieutord2016}. Typically, a model uses 12 spheroidal shells with 30 points in a Gauss-Lobatto grid radially and 24 points in a Gauss-Legendre grid in latitude. Time evolution is insured by a first order backward Euler method where the time step is manually set to 0.5\,Myr, and is decreased when the model fails to converge.

Following the theoretical work of \cite{Zahn1992} on the predicted diffusion constant resulting from rotationally-induced chemical mixing, and the one-dimensional implementation by \cite{Mombarg2022}, the vertical chemical diffusivity is taken as,

\begin{equation} \label{eq:Dv}
    D_{\rm v} = \eta \left< N_0^2\right>_V^{-1} \left< K r^2  \left(\vn\cdot\nabla \Omega\right)^2 \right>_\theta.
\end{equation}

Here, $K$ is the thermal diffusivity, $\vn$ is the unit vector normal to the isobars, and $\Omega$ is local angular velocity. The term $\left< N_0^2\right>_V$ is the volume-averaged squared Brunt-V\"ais\"al\"a frequency of the initial steady-state model, where the average is taken over the volume where $N_0^2 > 0$. We thus smooth out the rapid radial variations of $N^2$ that often raise numerical difficulties. Furthermore, $\eta$ is a free non-dimensional parameter, which we adjust to avoid numerical instabilities arising if chemical diffusion is too low, \edit{while keeping it close to unity. For model M1, $\eta$ needs to be increased to compensate for the smaller shear compared to model M2.}  The values of $\eta$ are given in Table~\ref{TS} of the appendix.
Furthermore, we take the angular average of \edit{the term between $\left< \cdot \right>_\theta$ in Eq.~(\ref{eq:Dv})} and assume a fixed horizontal diffusion coefficient $D_{\rm h} = 10^5\,{\rm cm^2\,s^{-1}}$. In the \ester evolution models, there is no ad-hoc enhanced mixing at the core boundary \edit{(overshooting)} as is typically applied in one-dimensional evolution models to account for the discrepancy between predicted and observationally-inferred core masses. 

\begin{figure}
    \centering
    \includegraphics[width = 0.85\columnwidth]{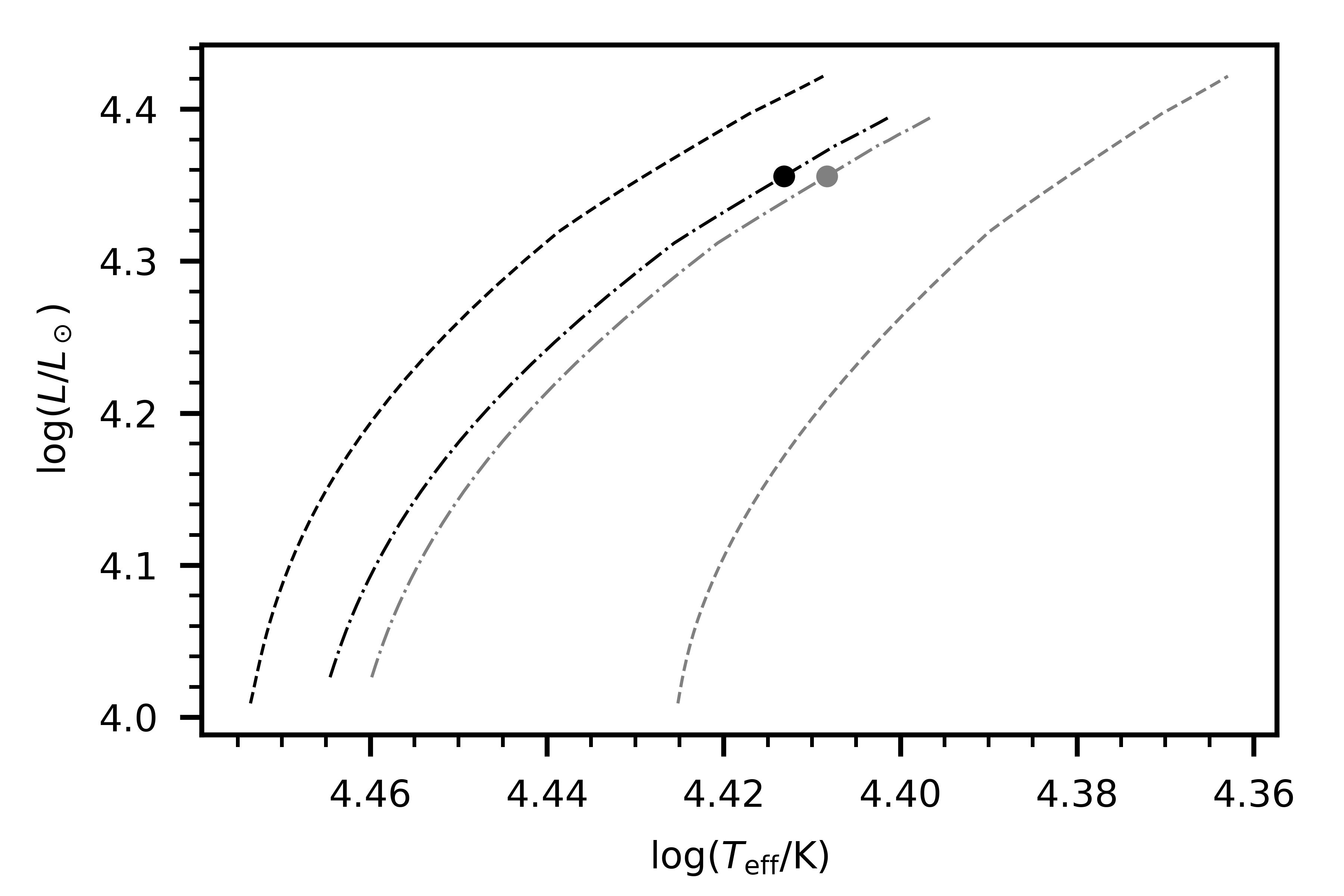}
    \caption{HRD showing the evolution tracks of the 2-D models for a 12\dMsun star ($Z = 0.012, X_{\rm i} = 0.71$) for \omegai = 0.15 (dashed-dotted), and 0.5 (dashed). The black lines represent the properties at the pole, the grey lines the properties at the equator. The tracks are terminated slightly before the terminal-age main sequence when the solver can no longer converge, which occurs at \xcx = 0.141, and 0.131 for the aforementioned rotation rates, respectively. The dots mark the model of HD192575 that is discussed in Sect.~\ref{sec:astero}.  }
    \label{fig:HRD}
\end{figure}

\begin{figure}
        \centering
    \includegraphics[width = 0.9\columnwidth]{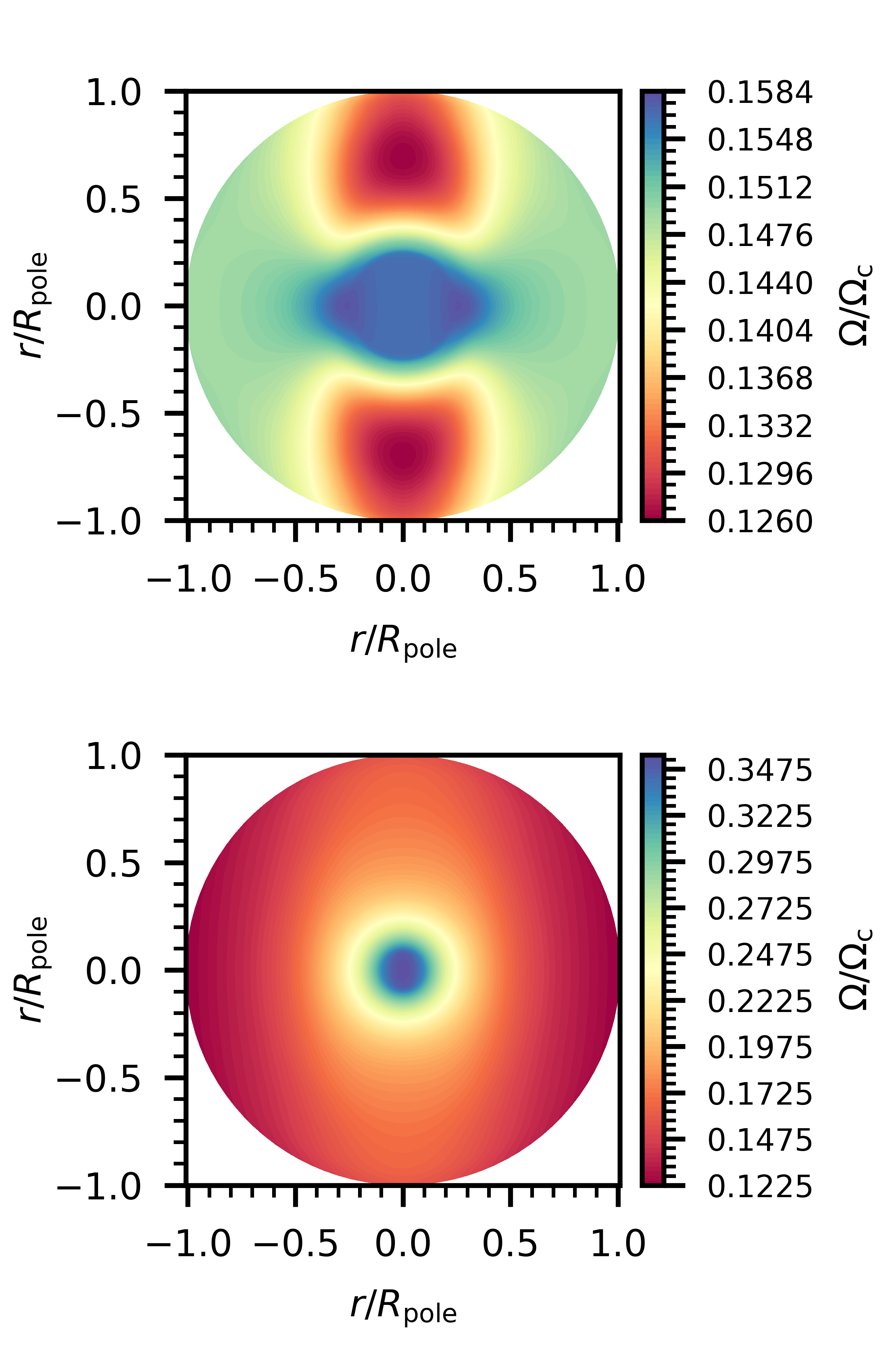}
    \caption{Map of the angular velocity as a fraction of the critical angular velocity for a model of 1\,Myr (top panel) and 15.75\,Myr (bottom panel). \edit{The plots show cuts in the meridian plane.}
    These plots are for a 12\dMsun star with \omegai = 0.15.    }
    \label{fig:omega_map}
\end{figure}

We have successfully ran two-dimensional evolution models for a 12\dMsun star, once for an initial rotation \edit{(at the equator)} of 15\percent the \edit{critical} angular velocity \edit{$\Omega_{\rm c}$} (model M1, \edit{$v_{\rm eq} = 112\,{\rm km\,s^{-1}}$}), and once for 50\percent (model M2, \edit{$v_{\rm eq} = 358\,{\rm km\,s^{-1}}$})\footnote{\edit{The models are available on \url{https://doi.org/10.5281/zenodo.8228904}.}}. The evolution models start from a steady-state model at the ZAMS with a uniform chemical composition with $X=0.71$ and $Z=0.012$ \edit{(no assumption of solid-body rotation)}. The evolution in the Hertzsprung-Russell diagram (HRD) of the models computed with \ester is shown in Fig.~\ref{fig:HRD} for the effective temperature at the pole and at the equator. Figure~\ref{fig:omega_map} shows the angular velocity distribution at 1~Myrs and 15.75~Myrs for model M1. This stellar model starts with equatorial regions rotating more rapidly than polar ones, namely solar-like, and gradually evolve towards a more shellular rotation with a slightly anti-solar surface rotation.
The same behaviour is also seen in M2 \edit{(see Appendix~\ref{ap:omega_M2})}. Contrary to an evolution modelled with a series of steady-state models, where one decreases the hydrogen mass fraction in the core \citep{Gagnier2019b}, our model evolution shows that $\Omega/\Omega_{\rm c}$ decreases as the star evolves. We understand this behaviour as a consequence of the slow redistribution of AM through baroclinic modes, which are damped on a time scale similar to the nuclear one. The time scale on which baroclinic modes are damped is given by
\begin{equation}
   \tau_{\rm baro} = \left< \frac{N^2}{\Omega^2 K}\right>_V \frac{d_{\rm env}^2}{(\pi^2 + 4)\pi^2},  
\end{equation}
following the work of \cite{busse81}. The $\tau_{\rm baro}$ is usually of the order of the Eddington-Sweet time scale \citep{R05}. In its expression, $d_{\rm env}$ is the thickness of the radiative envelope in the polar direction. The nuclear evolution time scale, $\tau_{\rm evol} = X_{\rm c}/\dot{X_{\rm c}}$, is found to be roughly 3 and 30 times larger at the start of the main sequence (MS) for M1 and M2 respectively. Hence, and especially for M2, baroclinic modes that may be excited by initial conditions can be damped during nuclear evolution. We can thus assume that initial conditions are of little importance and are forgotten during the first part of the main sequence. Yet, at the end of the MS the ratio of time scales is reversed: the damping of baroclinic modes happens on a time scale which is 100 (M1) and 10 (M2) times longer than the nuclear one. It implies that the dynamical evolution, and especially the rotation rates, cannot be computed as a succession of stationary states as in \cite{Gagnier2019b}. Table~\ref{TS} in Appendix~\ref{ap:timescales} lists the values of these time scales for our two models. We note that the growth of the stellar radius is the main contributor to the growth of $\tau_{\rm baro}$ with MS evolution.

\begin{figure}
    \centering
    \includegraphics[width = 0.9\columnwidth]{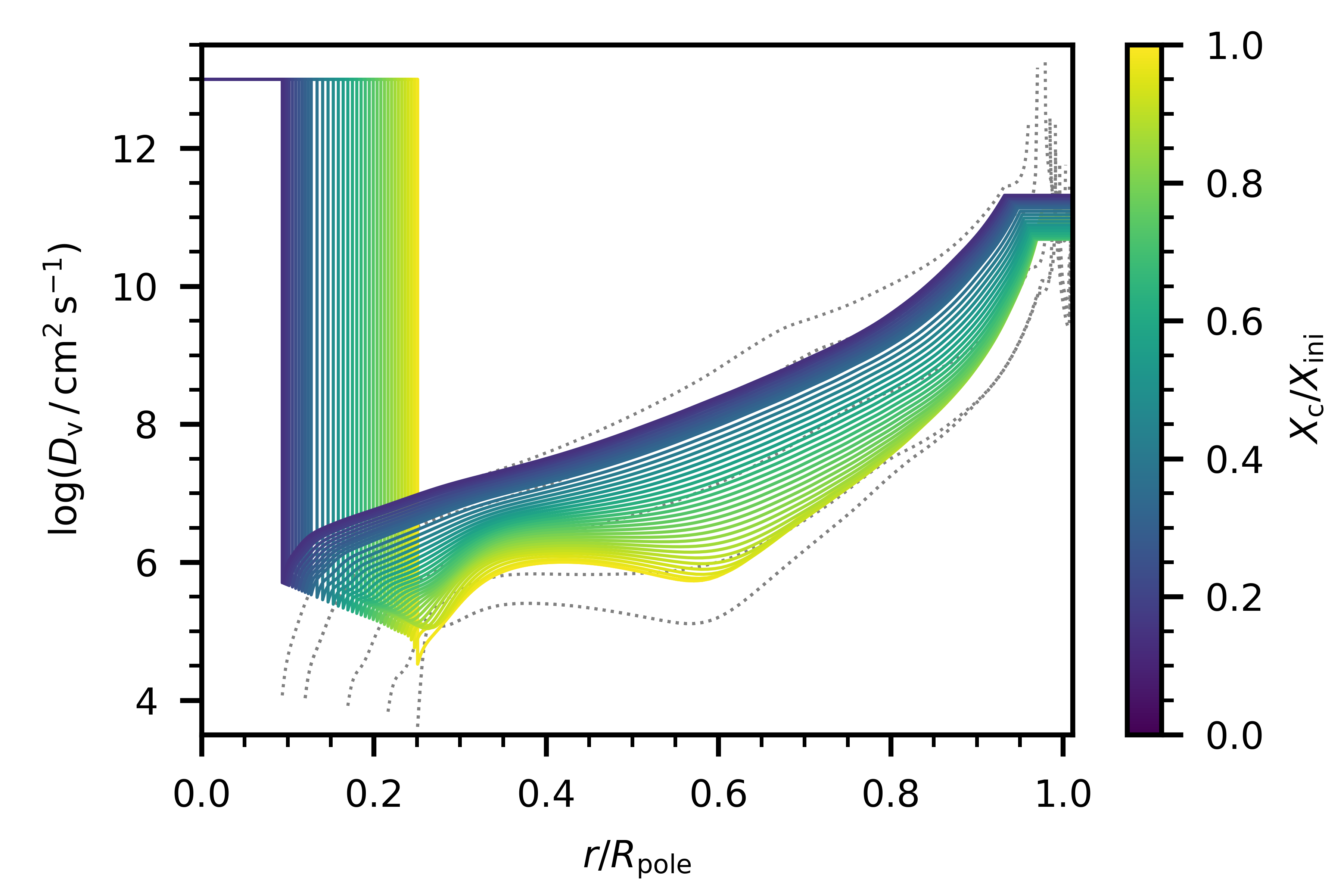}
    \caption{The profile of the vertical diffusion coefficient as per Eq.~(\ref{eq:Dv}) for a 12\dMsun model with an initial rotation frequency of 15\percent the \edit{critical} angular velocity. The grey dotted lines show $D_{\rm v}\left< N_0^2\right>_V/(\eta N^2)$ at a few ages \edit{(from left to right: \xcx = 0.14, 0.30, 0.58, 0.82, 0.98)}, that is, if we were to take the local value of the Brunt-V\"ais\"al\"a frequency instead of the volume averaged value.   }
    \label{fig:Dmix}
\end{figure}

Figure~\ref{fig:Dmix} shows the evolution of the vertical diffusion coefficient as described by Eq.~(\ref{eq:Dv}) for model M1. At the start of the main sequence the latitudinally average angular velocity $\overline{\Omega}(r)$ decreases from the core towards the surface until roughly 60\percent of the fractional radius, after which $\overline{\Omega}$ increases towards the surface \edit{(see right panel of Fig.~\ref{fig:omega_prof})}, creating a layer where the shear is weak and thus $D_{\rm v}$ reaches a local minimum. Further along the main-sequence evolution, $\overline{\Omega}$ keeps decreasing all the way towards the surface. Figure~\ref{fig:Dmix} also shows the difference between taking an average value for $N^2$ as we do, compared to using the full profile. This latter case shows the rapid variations of $D$ that occur near the core and which raise numerical problems as mentioned above.

\section{Calibration to HD192575} \label{sec:astero}

\begin{figure*}[t]
    \centering
    \includegraphics[width = 0.75\textwidth]{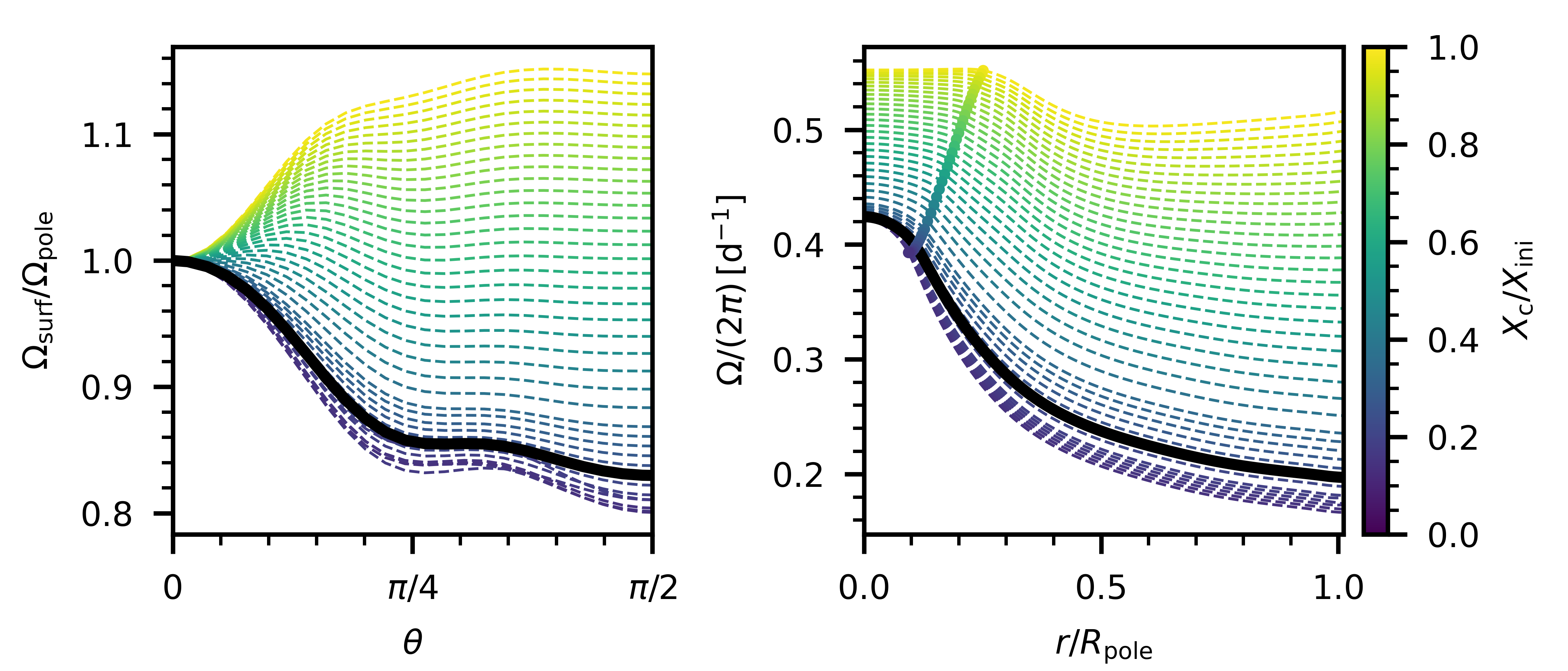}
    \caption{Left panel: Predicted surface rotation as a function of colatitude throughout the evolution (indicated by the colour), normalised by the rotation frequency at the pole. Right panel: Profile of the angular velocity (averaged over $\theta$) throughout the evolution. The thick coloured line indicates the location of the convective core boundary. The black lines in both panels correspond to the best-matching model for HD192575 (last column in Table~\ref{tab:parameters}).  }
    \label{fig:omega_prof}
\end{figure*}

To calibrate our rotating models, we use the results of \cite{Burssens2023}, hereafter B23, who have performed an asteroseismic modelling of the $\beta$~Cephei pulsator HD192575, based on one-dimensional non-rotating stellar evolution models. This star is a unique case to test the theory of angular momentum transport, as it has a relatively precise age-estimate, and an inferred rotation profile from the rotational splittings of the observed multiplets. The stellar parameters of HD192575 derived by \citetalias{Burssens2023} are summarised in the middle column of Table~\ref{tab:parameters}. We have computed \ester evolution tracks with a mass ($M_\star$), metallicity ($Z$) and initial hydrogen-mass fraction ($X$) fixed to the values found by \citetalias{Burssens2023}. We then picked the model for which, after evolution, the hydrogen-mass fraction in the convective core is closest to that inferred by these authors. 
Since the shape of the internal rotation profiles in massive stars is currently unknown, these authors modelled the rotational splittings by assuming a rotation profile that is described as,   
\begin{equation} \label{eq:rotprof}
\Omega(r) = \left\{
        \begin{array}{ll}
            \Omega_{\rm core} & \quad r \leq r_{\rm core}, \\
            \Omega_{\rm core} - \Delta \Omega \frac{r -r_{\rm core}}{r_{\rm shear} - r_{\rm core}} & \quad r_{\rm core} < r < r_{\rm shear}, \\

            \Omega_{\rm surf} & \quad r \geq r_{\rm shear}, \\
        \end{array}
    \right.
\end{equation}
where $\Delta \Omega = \Omega_{\rm core} - \Omega_{\rm surf}$, $r_{\rm core}$ the radius of the convective core, and $r_{\rm shear}$ the outer radius of the shear zone where the rotation frequency decreases linearly until it reaches the surface value. The values of $\Omega_{\rm core}$ and $\Omega_{\rm surf}$ are then optimised to best reproduce the observations for a given assumption for $r_{\rm shear}$. In the \ester models, however, the rotation profile is computed self-consistently. Figure~\ref{fig:omega_prof} shows the predicted evolution of the rotation profile for model M1. As can be seen from this figure (left), the latitudinal differential rotation changes from solar-like to anti-solar. This change may be interpreted as follows. The solar-like latitudinal profile is indeed the relaxed steady baroclinic state of a rotating radiative envelope \citep{Espinosa2013}. As time evolution proceeds, the core shrinks and spins up due to angular momentum conservation. Thanks to the Taylor-Proudman theorem, which states that the velocity field cannot vary \edit{in the direction of} the rotation axis\footnote{This is exactly true for a steady solution of an incompressible inviscid rotating fluid when the Coriolis term dominates all other terms \citep{rieutord15}, but it can be extended to fluids of varying density if momentum $\rho\vv$ is used instead of $\vv$.}, we understand that polar regions tend to follow the core and thus turn out to rotate more rapidly than equatorial regions. Obviously this is a consequence of the rapid nuclear evolution which prevents the star to relax to a quasi-steady state.

Figure~\ref{fig:omega_prof} (right panel) also shows the ``radial" rotation profile averaged over the latitude. It looks similar to the linear piece wise profile given in  Eq.~(\ref{eq:rotprof}) when $r_{\rm shear}$ is taken equal to the outer radius of the region with a non-zero gradient in the mean molecular weight ($\mu$). This behaviour seems to be independent of the initial rotation since model M2 shows the same profile. Therefore, we compare our results with the core and surface rotation frequencies derived with this assumption. The value of $\Omega_{\rm core}$ in the \ester models is defined as the value of $\overline{\Omega}(r_{\rm core})$, where an average is taken over all points in latitude.
The last column in Table~\ref{tab:parameters} shows the parameters of the \ester M1 - model. The predicted core- and surface rotation of this model are consistent with the observationally derived values for HD192575. Moreover, this model is able to reproduce, at the inferred age, the core mass, asteroseismic radius from the 1-D modelling, the astrometric luminosity from Gaia \citep{Gaia2022}, and the spectroscopically-derived effective temperature within the uncertainties quoted in \citetalias{Burssens2023}.
It should be noted that with the implemented rotational mixing, the age is consistent with the age predicted from 1-D \edit{(MESA)} models with core-boundary mixing, although its efficiency could not be precisely constrained by \citetalias{Burssens2023}. 
In summary, the 2-D evolution model M1 is able to explain the measured core and surface rotation of HD192575 at the asteroseismically inferred age. \newline

\begin{figure}[htb]
    \centering
    \includegraphics[width = 0.87\columnwidth]{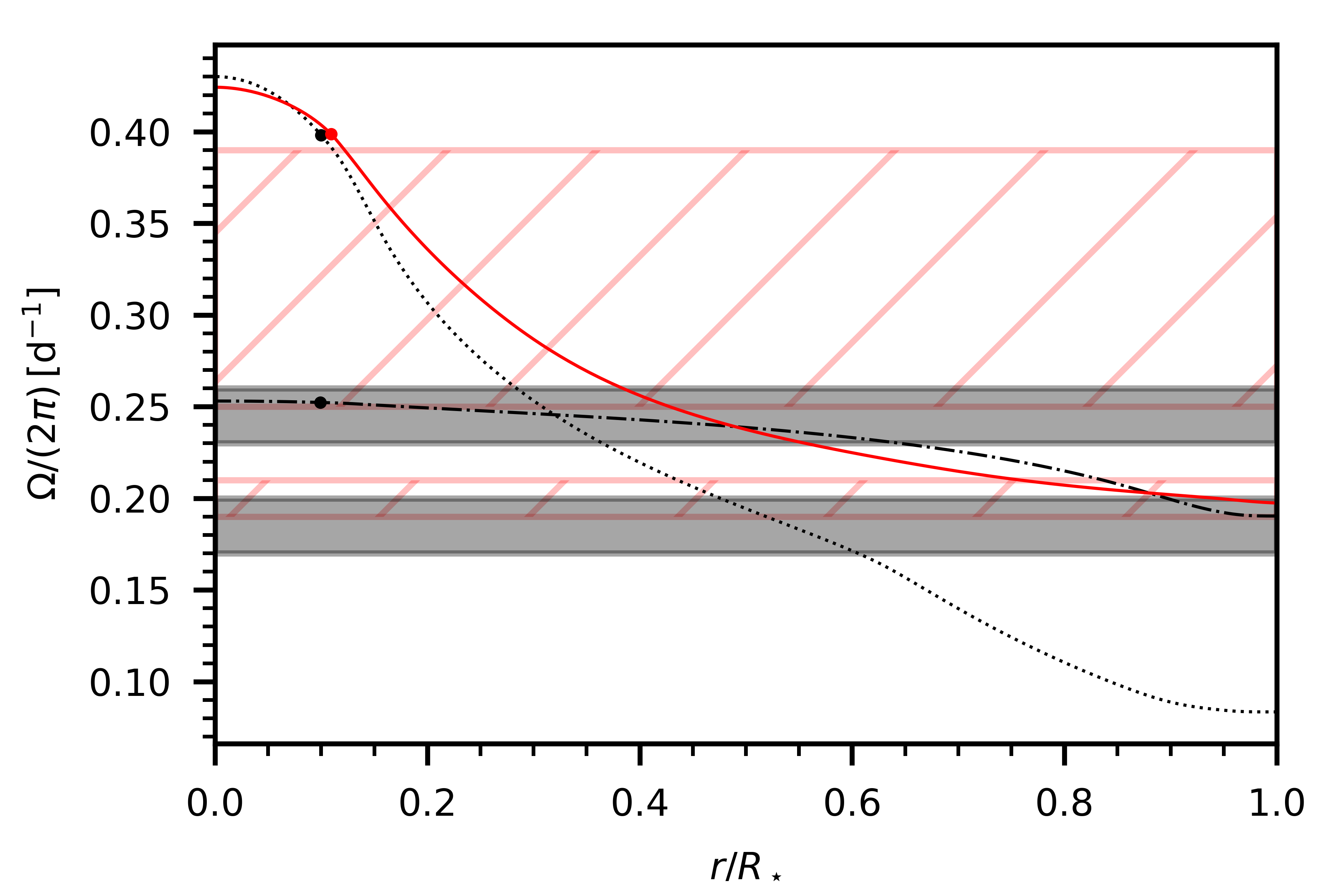}
    \caption{Predicted rotation profile of HD192575 with \ester (red solid line) and \mesa (black lines). The profiles indicated by a solid line and dotted line are for a constant viscosity of $10^7\,{\rm cm^2\,s^{-1}}$, the dashed line is for $10^9\,{\rm cm^2\,s^{-1}}$. The dots correspond to the location of the core boundary. The red and grey shaded areas indicate the measured core (top one) and surface (bottom one) rotation frequencies by \cite{Burssens2023}, assuming $r_{\rm shear}$ equal to the outer boundary of $\mu$-gradient zone, and $r_{\rm shear} = R_\star$, respectively.} 
    \label{fig:mesa_vs_ester}
\end{figure}

Additionally, we have computed 1-D SSE models with \mesa \citep[r22.11.1;][]{Paxton2011, Paxton2013, Paxton2015, Paxton2018, Paxton2019, Jermyn2023}, using the same physics as \citetalias{Burssens2023}, except that we also include rotation in our models \edit{(more details in Appendix~\ref{ap:mesa})}. The rotation profiles predicted by the 1-D \mesa models is similar to taking $r_{\rm shear} = R_\star$ (Eq.~(\ref{eq:rotprof})). Therefore, we compare the \mesa profiles with the core and surface rotation frequencies derived by \citetalias{Burssens2023} with this assumption. When the same uniform viscosity is assumed as the one used in the \ester models, we find that the 1-D \mesa model for \omegai = 0.15 does not match the core and surface rotation rates. To reproduce the measured core and surface rotation frequencies at the current age with the 1-D \mesa model, the viscosity has to be increased to $10^9\,{\rm cm^2\,s^{-1}}$ compared to the value of $10^7\,{\rm cm^2\,s^{-1}}$ used in the \ester models (vertical and horizontal components). While both the 1-D \mesa and 2-D \ester models can reproduce the core and surface frequencies of HD192575, the shape of the rotation profile is significantly different between the two (see Fig.~\ref{fig:mesa_vs_ester}). Future studies on asteroseismic rotation inversion might be able to rule out one of these profiles \citep{Vanlaer2023}.

\section{Nitrogen enhancement at the surface} \label{sec:mixing}

The observed abundance of $^{14}$N at the surface is often used to probe the efficiency of (rotational) chemical mixing in massive stars \citep[e.g.][]{Brott2011}.
The nuclear reactions in the \ester models are described by analytical formulae for the energy generation rates of the \textit{pp}-chain and CNO cycle \citep{Kippenhahn1990, Rieutord2016}. To track the abundance of $^{14}$N as a result of hydrogen-burning via the CNO cycle, we make the following assumptions.
First, at the start of the evolution all $^{12}$C present in the core is converted to $^{14}$N. Then, the reaction that controls the rate at which $^{14}$N is generated is the proton capture of $^{16}$O to create $^{17}$F, which then rapidly decays into $^{17}$O, itself reacting on protons to yield $^{14}$N and $^4$He. Therefore, the evolution of the mass fraction of $^{14}$N due to nuclear reactions is described by,

\begin{equation}
    \rho\frac{{\rm d}X(^{14}{\rm N})}{{\rm d}t} = \lambda(^{16}{\rm O} \rightarrow ^{17}{\rm F}) n(^{16}{\rm O}) n({\rm H}) m({\rm ^{14}N}),
\end{equation}
where $n(\cdot)$ denote the number densities, \edit{$m$ the atomic mass}, and $\lambda(^{16}{\rm O} \rightarrow ^{17}{\rm F})$ is the maxwellian-average reaction rate $\left<\sigma V\right>$, taken from \cite{Angulo1999},

\begin{equation}
    \calN_{\rm A}\lambda(^{16}{\rm O} \rightarrow ^{17}{\rm F}) = 7.37 \cdot 10^7e^{-16.696 T_9^{-1/3}} T_9^{-0.82},
\end{equation}
where $\mathcal{N}_{\rm A}$ is Avogadro's number. With this simple modelling we can reproduce the evolution of $^{14}$N core abundance as calculated through a more realistic network of nuclear reactions like in the MESA code.

In 1-D stellar evolution codes, the effect of chemical transport via meridional circulation  is typically treated in a diffusive way by adding an extra term to the vertical diffusion coefficient. This additional term takes on the following form $D_{\rm eff}~=~|ru_{\rm v}|^2/(30 D_{\rm h})$ \citep{Chaboyer1992},
where $u_{\rm v}$ is the vertical component of the meridional flow, and $D_{\rm h}$ the horizontal diffusion coefficient \citep[see Eq.~(18) in ][]{Palacios2003}. 

In the present work, using 2D-models, we tested the contribution of the meridional circulation to the transport of chemicals. Typical values for the maximum of $u_{\rm v}$ obtained from the 2-D models range from $10^{-4}$ to $3 \cdot 10^{-3}\,{\rm cm\,s^{-1}}$, which turn out to be too small to have any noticeable influence on element transport. As a result
surface nitrogen abundance for the \ester model of HD192575 is weak: $\Delta [{\rm N/H}] < 0.05$\,dex compared to the initial value.

In Appendix~\ref{ap:stream} we show the stream function of the meridional velocity field for a star at the start of the MS, and for a star near the end. As the rotation profile gradually evolves to a more shellular configuration, the angular dependency of the meridional circulation is mostly described by a spherical harmonic of degree $\ell= 2$. Thus, the two-dimensional models show that the assumption that higher order spherical harmonics can be neglected in the expansion of the meridional velocity field is not true in stars at the beginning of the MS, even at the low rotation rate of model M1.

\section{Conclusions} \label{sec:conclusions}

In this letter we presented the first results of a 2D-modelling of rotating stars including time evolution. These new models are an update of the 2D-\ester models designed by \cite{Espinosa2013} which compute steady 2D models of fast rotating stars. As the steady models, the time-dependent ones take into account the centrifugal flattening of the star as well as the large-scale flows (differential rotation and meridional circulation) driven by the baroclinicity of the star.

We ran two 12\dMsun models of a massive star with initial angular velocities of 15 and 50\percent of the critical one, respectively. The first model has a rather mild rotation rate but can be compared to the recent observation of a massive star, while the second model allowed us to test the performance of the code and revealed some new features of the internal dynamics of a massive star (see below).

The first model was actually designed to reproduce the observations of the $\beta$~Cephei pulsator HD192575 as derived by \cite{Burssens2023}. We found that our model gives a luminosity, effective temperature, and core mass that are consistent with the observationally derived values. Moreover, the rotation profile derived from the \ester model is also in accordance with the measurements of the core and envelope rotation rates of \cite{Burssens2023}. We note that these authors used a simplified rotation profile (e.g. Eq.~(\ref{eq:rotprof})) to derive the rotation rates. Fortunately, this profile turns out to be similar to the actual one predicted by the models, hence saving us from an inconsistent comparison.

Another result provided by the foregoing 2D-models is the weakness of the meridional circulation. In the dynamics of baroclinic flows this is controlled by viscosity to ensure the balance of angular momentum flux. The transport of chemicals by this flow seems to be negligible, but this need to be confirmed by a more detailed analysis since our models do not include the jump in viscosity expected at the core-envelope boundary, and which is expected to drive a Stewartson layer along the tangential cylinder of the core \citep{R06,GR20}.

Our 2D-models raise new questions on the dynamics of rotating stars. In particular, the possibility of using a succession of steady state models to monitor the rotational evolution of early-type stars, as done in \cite{Gagnier2019b}, is now questionable and needs new investigations. Our results indeed show that for the 12\dMsun model we computed, nuclear evolution is slow enough for relaxing the star to a quasi-steady state only at the beginning of the MS. When the star comes near the end of the MS the nuclear evolution becomes faster than the damping time of baroclinic modes \citep{busse81}. The question then arise as to when a succession of steady models is liable to represent the evolution of a rotating star. This is a complex question that is of course related to the general one of the angular momentum transport in stars still pending since the first \edit{measurements of differential rotation in red giant stars \cite[e.g.][]{beck2012,Deheuvels2012,deheuvels2015, mosser2012}}. New investigations with the present 2D-time-dependent \ester models will be presented in forthcoming articles.

\def\arraystretch{1.2}
\begin{table}[]
    \centering
    \caption{Comparison of \ester with the observations.}
    \begin{tabular}{ccc}
    \hline \hline
        Parameter & \cite{Burssens2023} & 2-D model   \\
        \hline
        $M_\star$ & $12 \pm 1.5$\,\Msun & 12\,\Msun  \\
        $Z$ & $0.012_{-0.000}^{+0.004}$   & 0.012  \\
        $X$ & 0.71  & 0.71  \\
        \xc & $0.176_{-0.045}^{+0.035}$ & 0.168  \\
        $f_{\rm CBM}$ & $0.030_{-0.025}^{+0.005}$ & - \\
        \hline
        $\tau$ & $17_{-5.4}^{+4.7}$\,Myr & 14.75\,Myr  \\
        $m_{\rm cc}$ & $2.9_{-0.8}^{+0.5}$\,\Msun  & 2.59\,\Msun \\
        $\log(L/{\rm L_\odot})$ & $4.30 \pm 0.07$ &  4.36 \\
        $T_{\rm eff, p}$ & \multirow{2}{*}{$23900 \pm 900$\,K} & 25892\,K\\
        $T_{\rm eff, e}$ &  & 25604\,K  \\
        
        $\Omega_{\rm core}/(2 \pi)$ & $0.30_{-0.05}^{+0.09}\,{\rm d^{-1}}$ & 0.40\,${\rm d^{-1}}$ \\
        
        $\Omega_{\rm surf}/(2 \pi)$ & $0.20 \pm 0.01 {\rm d^{-1}}$ & 0.20\,${\rm d^{-1}}$ \\
        
        $(\Omega/\Omega_{\rm c})_{\rm ini}$ & -  & 0.15 \\
        
        $R_{\rm \star, p}$ & \multirow{2}{*}{$9.1_{-1.7}^{+0.8}\,{\rm R_\odot}$}  & $7.57\,{\rm R_\odot}$ \\
        $R_{\rm \star, e}$ & & $7.64\,{\rm R_\odot}$ \\             
        \hline
    \end{tabular}
    \tablefoot{Stellar parameters of HD192575 derived by \cite{Burssens2023} and those of the \ester model for this same star. The parameters above the horizontal line are input for the model, while those below are output.}
    \label{tab:parameters}
\end{table}

\appendix 
\section{Stellar equations} \label{ap:eq}
In addition to Eq.~(\ref{eq:X_evol}), the \ester code solves the following set of fundamental stellar equations.
\begin{itemize}
\item Mass conservation
 \begin{equation}
 \dt{\rho}+\nabla\cdot(\rho \bold{v}) = 0, \label{eq:omega3}
 \end{equation}
where $\rho$ is the density and $\bold v$ the meridional velocity.

\item Meridional momentum equation
\begin{equation}
 \frac{1}{\rho}\nabla P + \nabla \phi - s \Omega^2 \bold{\hat{e}_s}  = \bold{F_{\rm visc}^{\rm merid}}, \label{eq:omega1}
 \end{equation}
 where $P$ is the pressure, $\phi$ the gravitational potential, $s$ the radial distance to the rotation axis, $\Omega$ the local angular velocity and $\bold{F}_{\rm visc}^{\rm merid}$ the meridional components of the viscous force. In this equation the time derivative of the meridional circulation has been neglected because of its extremely small value compared to other terms.

 \item Angular momentum equation
 \begin{equation}
\dt{s^2\Omega} + \bold{v} \cdot \nabla(s^2 \Omega) = \frac{1}{\rho} \nabla(\rho \nu s^2 \nabla \Omega), \label{eq:omega2}
 \end{equation}
 where $\nu$ is the kinematic viscosity. \edit{The kinematic viscosities in both the horizontal and vertical direction are free parameters in \ester. For both, we take $10^7\,{\rm cm^2\,s^{-1}}$, as this was found to be the order of magnitude needed to explain the rotation profiles of F-type stars by \cite{Mombarg2023}.}

 \item Entropy equation
  \begin{equation}
  \rho T \left( \dt{S}+\vv\cdot\nabla S\right) = \nabla\cdot (\chi\nabla T) + \rho\varepsilon,
  \end{equation}
  where $S$ is the entropy, $\chi$ the thermal conductivity, and $\varepsilon$ the energy generation rate per unit mass.

  \item Nuclear energy generation is computed with a simple law given by \cite{Kippenhahn1990}, namely  
  \begin{equation}
 \eps_*(\rho,T_9,X,Z) = \eps_0(X,Z)\rho T_9^{-2/3}\exp\left( -A/T_9^{1/3}\right) \left( 1 + C(T_9)\right),     
  \end{equation}
    from which we deduce the hydrogen consumption $\dot X_{\rm nuc}$. Here, $T_9 = T/10^9\,{\rm K}$, $A = 15.228$, and $C(T_9)$ is a correction term. 

\end{itemize}

\edit{
\section{Rotation profiles} \label{ap:omega_M2}
In this Appendix, we also show the rotation profiles of the M2 model (\omegai = 0.5).
\begin{figure}
    \centering
    \includegraphics[width = 0.9\columnwidth]{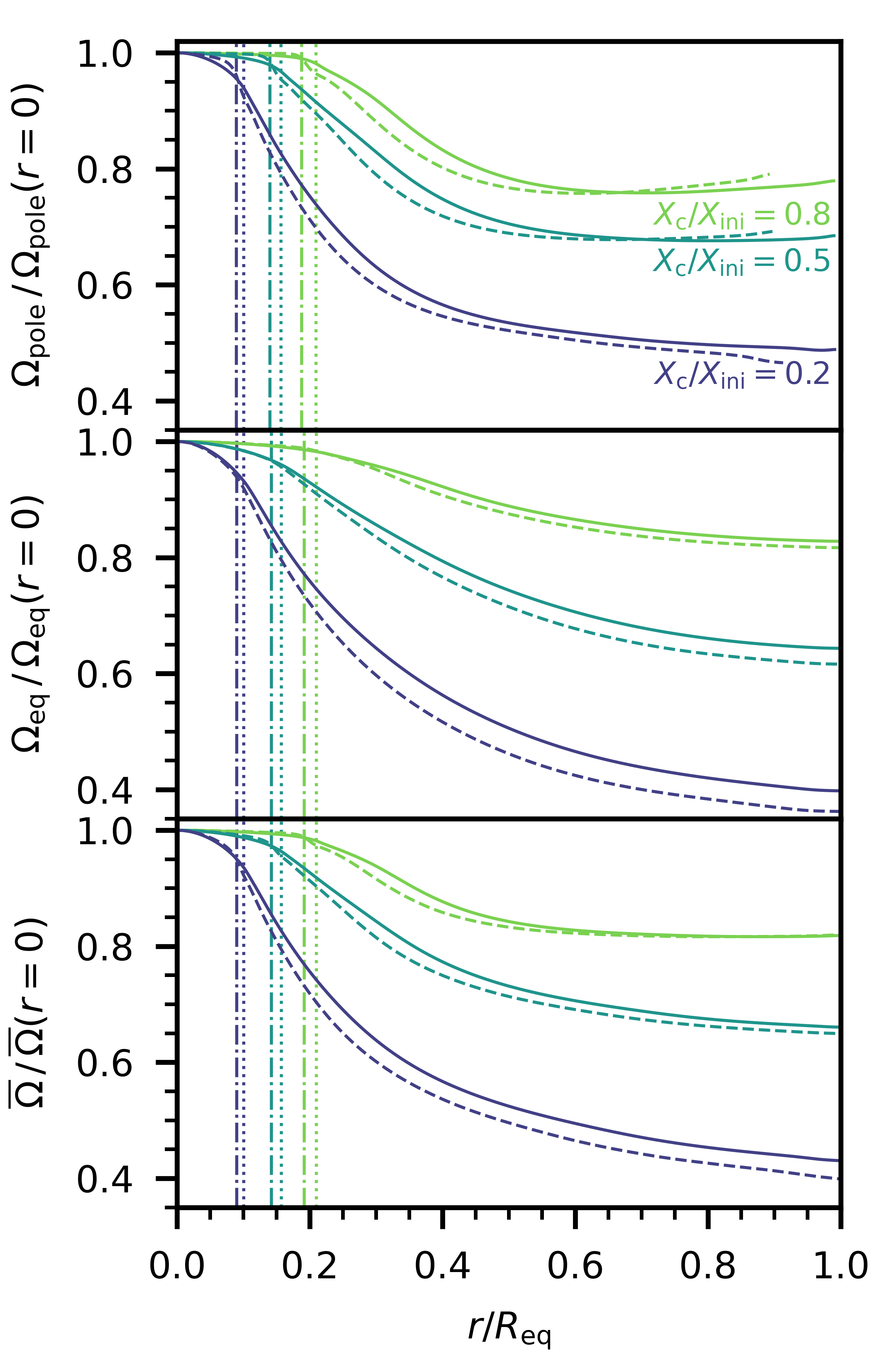}
    \caption{Profiles of the angular velocity as a function of the radial coordinate in the polar direction (top panel), equatorial direction (middle panel), and averaged over latitude (bottom panel). The solid lines correspond to model M1, the dashed lines to model M2. The locations of outer edge of the convective core are shown as vertical lines (dotted lines for model M1, dashed-dotted for model M2). }
    \label{fig:omega_M2}
\end{figure}

}

\section{Time scales} \label{ap:timescales}
In this Appendix, we show the nuclear and baroclinic time scales (see Sect.~\ref{sec:ester}), once close to zero-age main sequence (ZAMS), and once close to terminal-age main sequence (TAMS).
\begin{table}[htb]
    \centering
    \caption{Diffusivity factor and characteristic time scales of the models.}\label{TS}
    \begin{tabular}{cccccc}
    \hline     \hline

    Model    & $(\Omega/\Omega_{\rm c})_{\rm i}$ & $\eta$ & Age &$\tau_{\rm nuc}$ & $\tau_{\rm baro}$  \\
        &         &       &            &   (Myrs)       &   (Myrs)   \\ 
        \hline
     M1 & 0.15 & 9.48 & ZAMS & 2.72  &   0.95    \\ 
      &  &  &  TAMS &   0.12      &  7.55    \\ 
     M2 & 0.50 & 0.72 & ZAMS & 2.80  &   0.09 \\ 
        &  & &  TAMS &   0.10       & 0.80 \\ 
     \hline
    \end{tabular}
\end{table}

\edit{
\section{MESA setup} \label{ap:mesa}
In this Appendix, we provide a short summary of the physics used for the 1-D \mesa model of HD192575, discussed in Section~{\ref{sec:astero}}. The transport of AM in \mesa is treated as a diffusive process (Eq.~(B4) in \cite{Paxton2013}) and shellular rotation is imposed. As our aim here is to compare the 2-D \ester models with the 1-D physics used by \cite{Burssens2023}, we use the same description for the chemical mixing as these authors. This description is based on predictions of simulations of internal gravity waves \cite[e.g.][]{RogersMcElwaine2017, Varghese2023}, where the chemical diffusion coefficient takes the form of,
\begin{equation}
    D_{\rm IGW}(r) = D_0 \left( \frac{\rho_0}{\rho(r)} \right),
\end{equation}
where $D_0$ is a free parameter we set to $10^3\,{\rm cm^2\,s^{-1}}$ (same as \cite{Burssens2023}), and $\rho_0$ the density at the core boundary. This means that we set the factor $f_C$, that accounts for the different efficiencies between the transport of AM and chemical elements\footnote{\texttt{am\_D\_mix\_factor} in \mesa.} \citep{Heger2000}, equal to zero.
}

\section{Meridional circulation} \label{ap:stream}
In this Appendix, we show the stream lines of the meridional flow for models M1 (\omegai = 0.15) and model M2 (\omegai = 0.5).
\begin{figure}[htb]
        \centering
    \includegraphics[width = 0.8\columnwidth]{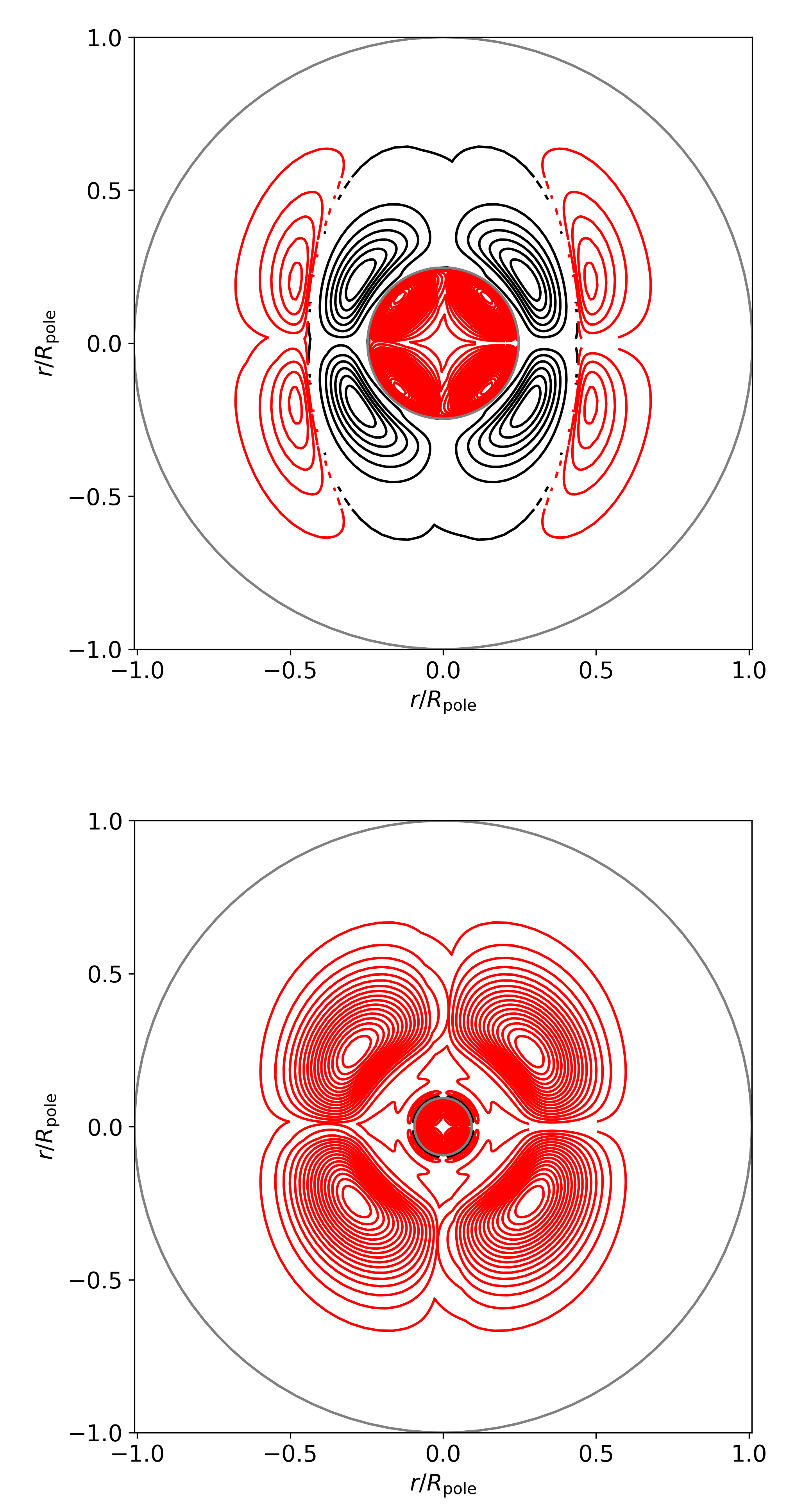}
    \caption{Isocontours of the stream function for model M1 of 1\,Myr (top panel) and 15.75\,Myr (bottom panel). \edit{The different colours indicate counter-rotating cells, where the red cells rotate clockwise in the first quadrant of the plot.} These models correspond to those shown in Fig.~\ref{fig:omega_map}. The isobars at the edge of the convective core and at the surface are shown in grey.  }
    \label{fig:stream15}
\end{figure}

\begin{figure}[htb]
        \centering
    \includegraphics[width = 0.8\columnwidth]{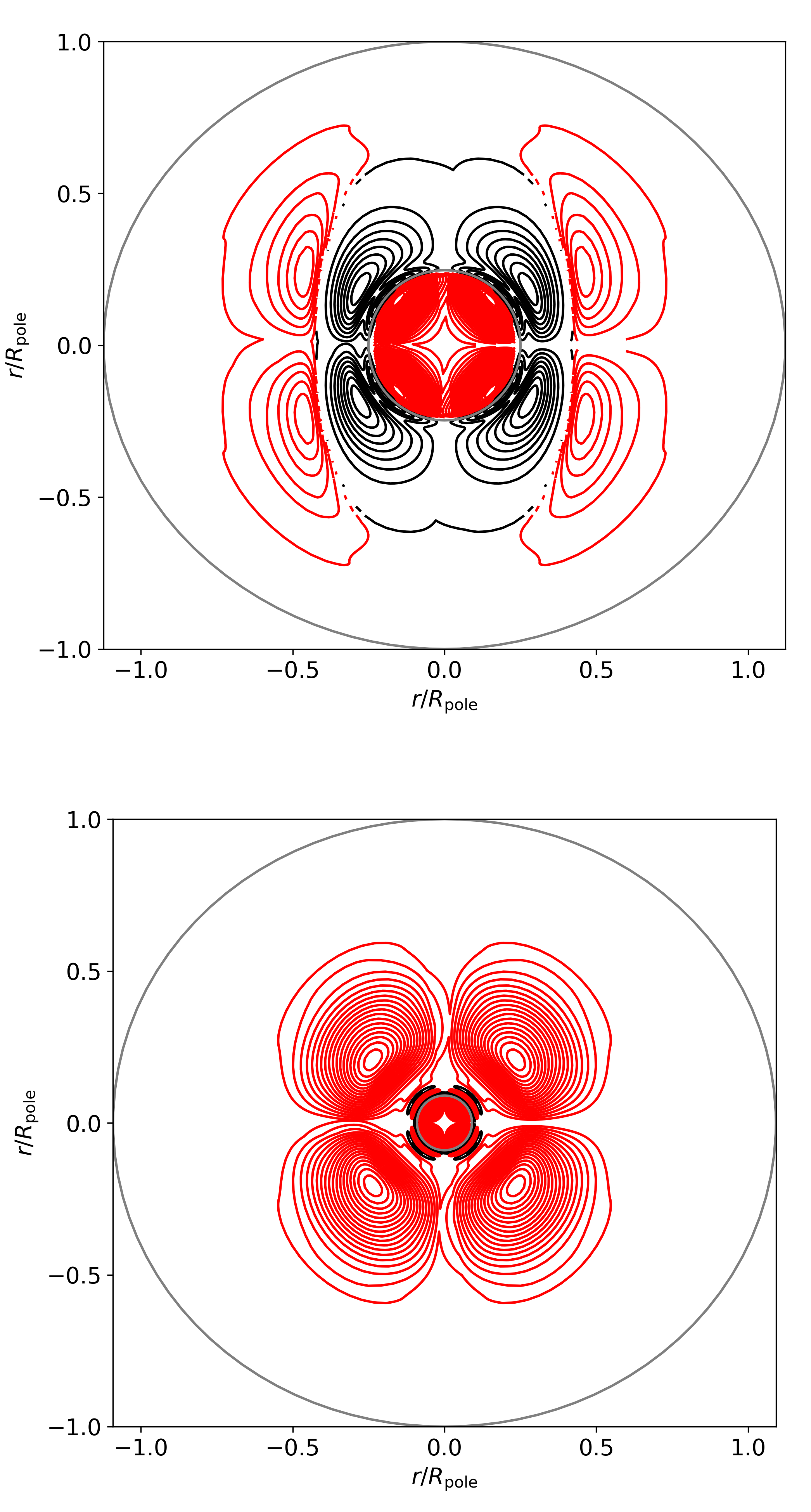}
    \caption{Same as Fig~\ref{fig:stream15}, but for model M2.}
    \label{fig:stream50}
\end{figure}

\begin{acknowledgements}
\edit{The authors are grateful to the referee prof. Georges Meynet for his comments and suggestions. }
 The research leading to these results has received funding the French Agence Nationale de la Recherche (ANR), under grant MASSIF (ANR-21-CE31-0018-02). The authors thank Siemen Burssens for providing the \mesa setup.
Computations of \ester 2D-models have been possible thanks to HPC resources from CALMIP
supercomputing center (Grant 2023-P0107).\end{acknowledgements}

\bibliographystyle{aa} 
\bibliography{main} 

%
%

\end{document}